\documentclass{llncs}
\usepackage[english]{babel}
\usepackage{amsmath,amssymb}
\usepackage{tikz}
\usepackage{graphicx}
\usepackage{makeidx}
\usepackage{multicol}
\usepackage{footmisc}
\usepackage{listings}
\lstMakeShortInline[basicstyle=\tt]|
\usepackage{lstomdoc}
\usepackage{paralist}
\usepackage{wrapfig}
\usepackage{stex-logo}
\usepackage[today,eso-foot]{svninfo}
\usepackage{url}

\usepackage[show]{ed}

\newcommand{\stexide}{\stex\kern-.5em I\kern-.1em D\kern-.3em E\xspace}
\newcommand{\LaTeXML}{\LaTeX{}ML\xspace}
\newcommand{\latexml}{\LaTeXML}
\newcommand{\omdoc}{OMDoc\xspace}

\newcommand{\texlipse}{\textsc{TeXlipse}\xspace}
\newcommand{\eclipse}{\textsc{Eclipse}\xspace}

\newcommand{\naive}{na\"\i ve\xspace}

\title{\protect\stexide: An Integrated Development Environment for \protect\stex Collections}
\author{Constantin Jucovschi \and Michael Kohlhase}
\institute{Computer Science, Jacobs University Bremen\\
\email{\{c.jucovschi,m.kohlhase\}@jacobs-university.de}}

\begin{document}
\svnInfo $Id: paper.tex 1152 2010-03-10 17:10:19Z kohlhase $
\svnKeyword $HeadURL:https://svn.kwarc.info/repos/supervision/mth/2010/jucovschi_constantin/papers/mkm.tex $
\maketitle

\begin{abstract}
\footnote{The final publication of this paper is available at \url{www.springerlink.com}}  Authoring documents in MKM formats like OMDoc is a very tedious task. After years of
  working on a semantically annotated corpus of \stex documents (GenCS), we identified a
  set of common, time-consuming subtasks, which can be supported in an integrated
  authoring environment.

  We have adapted the modular Eclipse IDE into \stexide, an authoring solution for
  enhancing productivity in contributing to \stex based corpora. \stexide supports
  context-aware command completion, module management, semantic macro retrieval, and
  theory graph navigation.
\end{abstract}
%\edexplanation

\svnInfo $Id: intro.tex 1240 2010-04-27 08:42:09Z cjucovschi $
\svnKeyword $HeadURL:https://svn.kwarc.info/repos/supervision/mth/2010/jucovschi_constantin/papers/mkm.tex $
\section{Introduction}\label{sec:intro}

Before we can manage mathematical `knowledge' --- i.e. reuse and restructure it, adapt its
presentation to new situations, semi-automatically prove conjectures, search it for
theorems applicable to a given problem, or conjecture representation theorems, we have to
convert informal knowledge into machine-oriented representations. How
exactly to support
this formalization process so that it becomes as effortless as possible is one of the main
unsolved problems of MKM. Currently most mathematical knowledge is available in the form
of {\LaTeX}-encoded documents. To tap this reservoir we have developed the
\stex~\cite{Kohlhase:ulsmf08,sTex:web} format, a variant of {\LaTeX} that is geared
towards marking up the semantic structure underlying a mathematical document.

In the last years, we have used \stex in two larger case studies. In the first one, the
second author has accumulated a large corpus of teaching materials, comprising more than
2,000 slides, about 800 homework problems, and hundreds of pages of course notes, all
written in {\stex}. The material covers a general first-year introduction to computer
science, graduate lectures on logics, and research talks on mathematical knowledge
management. The second case study consists of a corpus of semi-formal documents developed
in the course of a verification and SIL3-certification of a software module for safety
zone computations~\cite{KohKohLan:difcsmse10,KohKohLan:ssffld10}.  In both cases we took
advantage of the fact that \stex documents can be transformed into the XML-based
\omdoc~\cite{Kohlhase:OMDoc1.2} by the {\latexml} system~\cite{Miller:latexml},
see~\cite{KohKohLan:difcsmse10} and~\cite{DKLRZ:PubMathLectNotLinkedData10} for a
discussion on the MKM services afforded by this.

These case studies have confirmed that writing \stex is {\emph{much}} less tedious than
writing \omdoc directly. Particularly useful was the possibility of using the \stex-generated PDF
for proofreading the text part of documents. Nevertheless serious usability problems
remain. They come from three sources:
\begin{compactenum}
\item[\textbf{P}1] installation of the (relatively heavyweight) transformation system (with
  dependencies on \texttt{perl}, \texttt{libXML2}, {\LaTeX}, the \stex packages),
\item[\textbf{P}2] the fact that \stex supports an object-oriented style of writing mathematics, and 
\item[\textbf{P}3] the size of the collections which make it difficult to find reusable components.
\end{compactenum}
The documents in the first (educational) corpus were mainly authored directly in \stex via
a text editor (\texttt{emacs} with a simple \stex mode~\cite{Pesikan:cwcr06}). This was
serviceable for the author, who had a good recollection for names of semantic macros he had
declared, but presented a very steep learning curve for other authors (e.g. teaching
assistants) to join. The software engineering case study was a post-mortem formalization
of existing (informal) {\LaTeX} documents.  Here, installation problems and refactoring
existing {\LaTeX} markup into more semantic \stex markup presented the main problems. 

Similar authoring and source management problems are tackled by Integrated Development
Environments (IDEs) like \textsc{Eclipse}~\cite{Eclipse:web}, which integrate support for
finding reusable functions, refactoring, documentation, build management, and version
control into a convenient editing environment. In many ways, \stex shares more properties
with programming languages like \textsc{Java} than with conventional document formats, in
particular, with respect to the three problem sources mentioned above
\begin{compactenum}
\item[\textbf{S}1] both require a build step (compiling \textsc{Java} and formatting/transforming \stex
  into PDF/\omdoc),
\item[\textbf{S}2] both favor an object-oriented organization of materials, which allows to
\item[\textbf{S}3] build up large collections of re-usable components
\end{compactenum}

To take advantage of the solutions found for these problems by software engineering, we
have developed the \stexide integrated authoring environment for \stex-based
representations of mathematical knowledge. In the next section we recap the parts of \stex
needed to understand the system. In Section~\ref{sec:ui_features} we present the user
interface of the \stexide system, and in Section~\ref{sec:implementation} we discuss
implementation issues. Section~\ref{sec:concl} concludes the paper and discusses future
work.

%%% Local Variables: 
%%% mode: latex
%%% TeX-master: "paper"
%%% End: 

% LocalWords:  svnInfo kohlhase svnKeyword ednote stex ulsmf08 KohKohLan omdoc
% LocalWords:  difcsmse10 KohKohLan ssffld10 latexml latexml emph gencs textbf
% LocalWords:  PubMathLectNotLinkedData10 compactenum texttt texttt Pesikan
% LocalWords:  cwcr06 textsc stexide concl

\svnInfo $Id: stex.tex 1240 2010-04-27 08:42:09Z cjucovschi $
\svnKeyword $HeadURL:https://svn.kwarc.info/repos/supervision/mth/2010/jucovschi_constantin/papers/mkm.tex $
\section{\protect\stex: Object-Oriented {\LaTeX} Markup}\label{sec:stex}

The main concept in \stex is that of a ``{\emph{semantic macro}}'', i.e. a {\TeX} command
sequence $\mathcal{S}$ that represents a meaningful (mathematical) concept or object
$\mathcal{O}$: the {\TeX} formatter will expand $\mathcal{S}$ to the presentation of
$\mathcal{O}$. For instance, the command sequence |\positiveReals| is a semantic macro
that represents a mathematical symbol --- the set $\mathbb{R}^+$ of positive real
numbers. While the use of semantic macros is generally considered a good markup practice
for scientific documents\footnote{For example, because they allow adapting notation by macro
  redefinition and thus increase reusability.}, regular {\TeX/\LaTeX} does not offer any
infrastructural support for this. \stex does just this by adopting a semantic,
``object-oriented'' approach to semantic macros by grouping them into ``modules'', which
are linked by an ``imports'' relation.  To get a better intuition, consider the example in
listing~\ref{lst:stex-ex}.
\begin{lstlisting}[label=lst:stex-ex,caption=An \protect\stex module for Real Numbers,escapechar=|,language=sTeX]
\begin{module}[id=reals]
  \importmodule[../background/sets]{sets}
  \symdef{Reals}{\mathcal{R}}
  \symdef{greater}[2]{#1>#2}
  \symdef{positiveReals}{\Reals^+}
  \begin{definition}[id=posreals.def,title=Positive Real Numbers]
    The set $\positiveReals$ is the set of $\inset{x}\Reals$ such that $\greater{x}0$
  \end{definition}
  |\ldots|
\end{module}
\end{lstlisting}

which would be formatted to

\begin{quote}\hrule\vspace*{.3em}
  \textbf{Definition} 2.1 (Positive Real Numbers): \\ 
  The set $\mathbb{R}^+$ is the set of $x\in\mathbb{R}$ such that $x>0$\vspace*{.3em}\hrule
\end{quote}

Note that the markup in the module |reals| has access to semantic macro |\inset| (membership) from the
module |sets| that was imported by the document by |\importmodule| directive from
the \url{../background/sets.tex}. Furthermore, it has access to the |\defeq|
(definitional equality) that was in turn imported by the module |sets|. 

From this example we can already see an organizational advantage of \stex over {\LaTeX}:
we can define the (semantic) macros close to where the corresponding concepts are defined,
and we can (recursively) import mathematical modules. But the main advantage of markup in
\stex is that it can be transformed to XML via the {\latexml}
system~\cite{Miller:latexml}: Listing~\ref{lst:omdoc-ex} shows the
{\omdoc}~\cite{Kohlhase:OMDoc1.2} representation generated from the {\stex} sources in
listing~\ref{lst:stex-ex}.

\lstset{language=[1.3]OMDoc}
\begin{lstlisting}[label=lst:omdoc-ex,mathescape,escapeinside={\{\}},caption={An XML Version of Listing~\ref{lst:stex-ex}}]
<theory xml:id="reals">
 <imports from="../background/sets.{omdoc}#sets"/>
 <symbol xml:id="Reals"/>
 <notation>
   <prototype><OMS cd="reals" name="Reals"/></prototype>
    <rendering><m:mo>$\mathbb{R}$</m:mo></rendering>
 </notation>
  <symbol xml:id="greater"/><notation>$\ldots$</notation>
  <symbol xml:id="positiveReals"/><notation>$\ldots$</notation>
  <definition xml:id="posreals.def" for="positiveReals">
    <meta property="dc:title">Positive Real Numbers</meta>
    The set <OMOBJ><OMS cd="reals" name="postiveReals"/></OMOBJ> is the set $\ldots$
  </definition>
  $\ldots$
</theory>
\end{lstlisting}
One thing that stands out from the XML in this listing is that it incorporates all the
information from the \stex markup that was invisible in the PDF produced by formatting it
with {\TeX}.  

%%% Local Variables: 
%%% mode: latex
%%% TeX-master: "paper"
%%% End: 

% LocalWords:  svnInfo kohlhase svnKeyword stex emph mathcal mathbb lst symdef
% LocalWords:  lstlisting escapechar posreals.def ldots hrule vspace textbf
% LocalWords:  defeq latexml omdoc-ex omdoc lstset mathescape escapeinside
% LocalWords:  postiveReals

\svnInfo $Id: ui.tex 1240 2010-04-27 08:42:09Z cjucovschi $
\svnKeyword $HeadURL:https://svn.kwarc.info/repos/supervision/mth/2010/jucovschi_constantin/papers/mkm.tex $
\section{User interface features of \protect\stexide}\label{sec:ui_features}
One of the main priorities we set for \stexide is to have a relatively gentle learning curve.
As the first experience of using a program is running the installation process, we worked
hard to make this step as automated and platform independent as possible. We aim at 
supporting popular operating systems such as Windows and Unix based platforms (Ubuntu, SuSE).
Creating an OS independent distribution of Eclipse with our plugin preinstalled was a relatively 
straightforward task; so was distributing the plugin through an update site. What 
was challenging was getting the 3rd party software (|pdflatex|, |svn|, |latexml|, |perl|) and 
hence OS specific ports installed correctly. 

After installation we provide a new project wizard for \stex projects which lets the user
choose the output format (|.dvi|, |.pdf|, |.ps|, |.omdoc|, |.xhtml|) as well as one of the
predefined sequences of programs to be executed for the build process. This will control the
\eclipse-like workflow, where the chosen `outputs' are rebuilt after every save, and
syntactic (as well as semantic) error messages are parsed, cross-referenced, and displayed
to the user in a collapsible window.  The wizard then creates a stub project, i.e. a file
|main.tex| which has the structure of a typical {\stex} file but also includes |stex|
package and imports a sample module defined in |sample_mod.tex|.

\begin{figure}[ht]\centering\vspace*{-2em}
  \includegraphics[width=8cm]{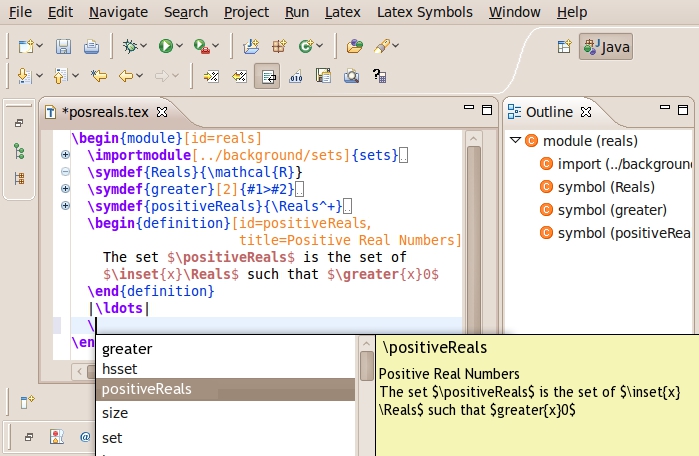}\vspace{-1em}
  \caption{Context aware autocompletion feature for semantic macros}\label{fig:autocomplete}\vspace*{-1em}
\end{figure}

\stexide supports the user in creating, editing and maintaining \stex documents or
corpora. For novice users we provide templates for creating modules, imports and
definitions. Later on, the user benefits from {\emph{context-aware autocompletion}}, which
assists the user in using valid {\LaTeX} and {\stex} macros. Here, by valid macros, we
mean macros which were previously defined or imported (both directly or indirectly) from
other modules. Consider the sample {\stex} source in listing \ref{lst:stex-ex}. At the end of
the first line, one would only be able to autocomplete {\LaTeX} macros, whereas at the end of
the second line, one would already have macros like |\inset| from the imported |sets| module
(see Fig. \ref{fig:autocomplete}). Note that we also make use of the semantic structure
of the \stex document in listing~\ref{lst:stex-ex} for explanations. Namely, the macro
|\positiveReals| is linked to its definition via the key
|for=positiveReals|, this makes it possible to display the text of the
definition as part of macro autocompletion explanation (the yellow box).

Similarly, {\emph{semantic macro retrieval}} (triggered by typing '|\*|') will suggest all
available macros from all modules of the current project. In case that
the auto-completed macro is not valid for the current context, \stexide will insert the required
import statement so that the macro becomes valid.

Moreover, \stexide supports several typical document/collection maintenance tasks:
Supporting {\emph{symbol and module name refactoring}} is very
important as doing it manually is both extremely error-prone and time
consuming, especially if two different modules define a symbol with the same
  name and only one of them is to be renamed. 
The {\emph{module splitting}} feature makes it easier for users to split a larger module intro
several semantically self contained modules which are easier to be reused.  This feature ensures
  that imports required to make the newly created module valid
  are automatically inserted.

\begin{wrapfigure}r{1.6cm}\vspace*{-2.5em}
    \begin{tikzpicture}[xscale=.5]
      \node (c) at (0,1) {C};
      \node (b) at (2,1) {B};
      \node (a) at (1,0) {A};
      \draw[->,dashed] (a) -- (c);
      \draw[->] (a) -- (b);
      \draw[->,dotted] (b) -- (c);
   \end{tikzpicture}\vspace*{-2em}
\end{wrapfigure}
At last, {\emph{import minimization}} creates warnings for unused or redundant
|\importmodule| declarations and suggests removing them. Consider for instance the
situation on the right, where modules |C| and |B| import module |A|. Now, if we add a
semantic macro in |C| that needs an import from |B|, then we should replace the import of
|A| in |C| with one of |B| instead of just adding the latter (i.e. we would replace the
dashed by the dotted import).

\begin{figure}[ht]\centering
  \includegraphics[width=10cm]{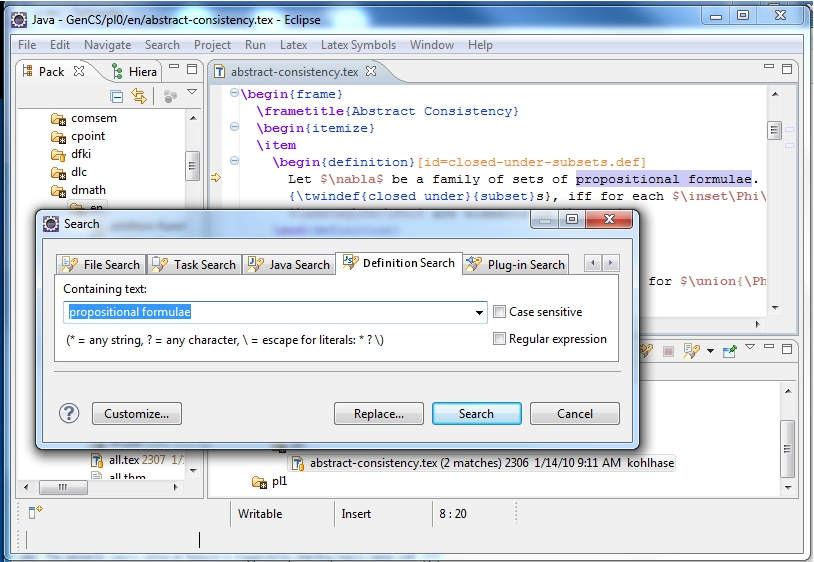}\vspace{-1em}
  \caption{Macro Retrieval via Mathematical Concepts}\label{fig:search}\vspace*{-1em}
\end{figure}

Three additional features make navigation and information retrieval in big corpora
easier. {\emph{Outline view}} of the document (right side of figure \ref{fig:autocomplete})
displays main semantic structures inside the current document.  One can use outline tree layout to copy, cut and
navigate to areas represented by the respective structures. In case of imports one can
navigate to imported modules. \emph{Theory graph navigation} is another feature of \stexide. It creates a
graphical representation of how modules are related through imports. This gives the author a
chance to get a better intuition for how concepts and modules are related. The last
feature is the \emph{semantic macro search feature}. The aim of this feature is to search
for semantic macros by their mathematical descriptions, which can be
entered into the search
box in figure \ref{fig:search}. The feature then searches definitions, assumptions and theorems
for the query terms and reports any |\symdef|-defined semantic macros `near' the
hits. This has proved very convenient in situations where the macro names are
abbreviated (e.g. |\sconcjuxt| for ``string concatenation by juxtaposition'') or if there
are more than one name for a mathematical context (e.g. ``concatenation'' for
|\sconcjuxt|) and the author wants to re-use semantic macros defined by someone else.

%%% Local Variables: 
%%% mode: latex
%%% TeX-master: "paper"
%%% End: 

% LocalWords:  svnInfo kohlhase svnKeyword stexide pdflatex svn latexml stex
% LocalWords:  omdoc xhtml ednote lst vspace includegraphics complition

\svnInfo $Id: arch.tex 1240 2010-04-27 08:42:09Z cjucovschi $
\svnKeyword $HeadURL:https://svn.kwarc.info/repos/supervision/mth/2010/jucovschi_constantin/papers/mkm.tex $
\section{Implementation}\label{sec:implementation}

\begin{figure}[ht]\centering\vspace*{-2em}
  \includegraphics[width=\linewidth]{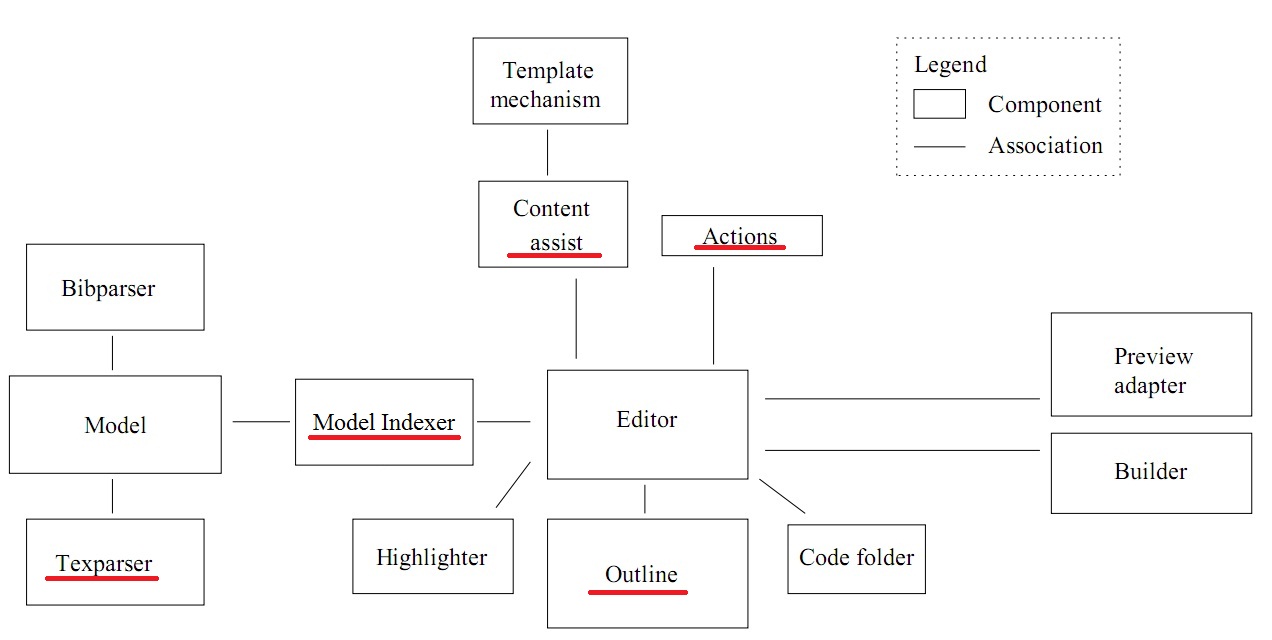}\vspace{-1em}
  \caption{Component architecture of \texlipse (adapted from
    \cite{TeXlipse:TechnicalSpec})}\label{fig:architecture1}\vspace*{-1em}
\end{figure}

The implementation of \stexide is based on the \texlipse \cite{TeXlipse:web} plugin for
Eclipse. This plugin makes use of \eclipse's modular framework (see
Fig. ~3
\ref{fig:architecture1}) and provides features like syntax highlighting, code folding,
outline generation, autocompletion and templating mechanisms. Unfortunately, \texlipse
uses a parser which is hardwired for a fixed set of {\LaTeX} macros like |\section|, |\input|, etc.  
which made it quite challenging to generalize it to \stex specific macros.
Therefore we had to reimplement parts of \texlipse so that \stex macros like |\symdef| and
|\importmodule| that extend the set of available macros can be treated specially. We have
underlined all the parts of \texlipse we had to extend or replace in
Figure~\ref{fig:architecture1}.

To support context sensitive autocompletion and refactoring we need to know the exact
position in the source code where modules and symbols are defined. Running a fully featured
{\LaTeX} parser like {\LaTeXML} proved to be too slow (sometimes
taking 5-10 sec to compile a document of 15 pages) and
sensitive to errors. For these reasons, we implemented a very fast but {\naive} {\LaTeX}
parser which analyses the source code and identifies commands, their arguments and
options. We call this parser {\naive} because it parses only one file a time
(i.e. inclusions, and styles are not processed) and macros are not expanded. We realize the
parse tree as an in-memory XML DOM to achieve format independence (see below). Then we run
a set of semantic spotters which identify constructs like module and import declarations,
inclusions as well as sections/subsections etc, resulting in an index of relevant
structural parts of the \stex source identified by unique URIs and line/column number
ranges in the source. For example, a module definition in \stex begins with
|\begin{module}[id=module_id]| and ends in a |\end{module}|, so the structure identifying
a module will contain these two ranges. 

Note that the {\LaTeX} document model (and thus that of \stex) is a tree, so two spotted
structure domains are either disjoint or one contains the other, so we implement a range
tree we use for efficient change management: \stexide implements a class which listens to
changes made in documents, checks if they intersect with the important ranges of the
spotted structures or if they introduce new commands (i.e.  start with
'|\|'). If this does not hold, the range tree is merely updated by calculating new line and column
numbers. Otherwise we run the {\naive} {\LaTeX} parser and the spotters again.

Our parser is entirely generated by a JavaCC grammar. It supports error recovery (essential
for autocompletion) and does not need to be changed if a new macro needs to be handled:
Semantic Spotters can be implemented as XQueries, and our parser architecture provides an
API for adding custom made semantic spotters. This makes the parser extensible to new \stex
features and allows working around the limitation of the {\naive} {\LaTeX} parser of not
expanding macros.
 
We implemented several indexes to support features mentioned in section
\ref{sec:ui_features}. For theory navigation we have an index called |TheoryIndex| which
manages a directed graph of modules and import relationships among them. It allows
\begin{inparaenum}[\em a)]
\item retrieving a list of modules which import/are imported by module $X$
\item checking if module $X$ is directly/indirectly imported by module $Y$.
\end{inparaenum}
|SymdefIndex| is another index which stores pairs of module URIs and symbols defined in
those modules. It supports fast retrieving of (symbol, module) pairs
where a symbol name
starts with a certain prefix by using a trie data structure. As expected, this index is used
for both context aware autocompletion as well as semantic macro retrieval features. The
difference is that context aware autocompletion feature also filters the modules not
accessible from current module by using the |TheoryIndex|.  Refactoring makes use of an
index called |RefIndex|. This index stores (module URI, definition module URI, symbol
name) triples for all symbol occurrences (not just definitions as in
|SymdefIndex|). Hence
when the author wants to rename a certain symbol, we first identify where that symbol
is defined (i.e. its definition module URI) and then query for all other symbols with same 
name and definition module URI.

%%% Local Variables: 
%%% mode: latex
%%% TeX-master: "paper"
%%% End: 

% LocalWords:  svnInfo kohlhase svnKeyword includegraphics linewidth texlipse
% LocalWords:  texipseSpec stexide templating stex symdef omdoc ednote lf xhtml
% LocalWords:  xspace xsltproc vspace inparaenum

\svnInfo $Id: concl.tex 1240 2010-04-27 08:42:09Z cjucovschi $
\svnKeyword $HeadURL:https://svn.kwarc.info/repos/supervision/mth/2010/jucovschi_constantin/papers/mkm.tex $
\section{Conclusion and Future Work}\label{sec:concl}
We have presented the \stexide system, an integrated authoring environment for \stex
collections realized as a plugin to the \textsc{Eclipse} IDE. Even though the
implementation is still in a relatively early state, this experiment confirmed the initial
expectation that the installation, navigation, and build support features contributed by
\textsc{Eclipse} can be adapted to a useful authoring environment for \stex with
relatively little effort. The modularity framework of \textsc{Eclipse} and the \texlipse
plugin for {\LaTeX} editing have been beneficial for our development. However, we were
rather surprised to see that a large part of the support infra\-structure we would have
expected to be realized in the framework were indeed hard-coded into the plugins. This has
resulted in un-necessary re-implementation work.

In particular, system- and collection-level features of \stexide like automated
installation, PDF/XML build support, and context-sensitive completion of command
sequences, import minimziation, navigation, and concept-based search have proven useful,
and are not offered by document-oriented editing solutions. Indeed such features are very
important for editing and maintaining any MKM representations. Therefore we plan to extend
\stexide to a general ``MKM IDE'', which supports more MKM formats and their
human-oriented front-end syntaxes (just like \stex serves a front-end to \omdoc in
\stexide).

The modular structure of \textsc{Eclipse} also allows us to integrate MKM services
(e.g. information retrieval from the background collection or integration of external
proof engines for formal parts~\cite{Lueth:PGEclipse}; see~\cite{KohRabZho:tmlmrsca10} for
others) into this envisioned ``MKM IDE'', so that it becomes a ``rich collection client''
to {\emph{a universal digital mathematics library (UDML)}}, which {\emph{would
    continuously grow and in time would contain essentially all mathematical knowledge}}
envisioned as the Grand Challenge for MKM in~\cite{Farmer:mkm05}.

In the implementation effort we tried to abstract from the \stex surface syntax, so that
we anticipate that we will be able to directly re-use our spotters or adapt them for other
surface formats that share the \omdoc data model. The next target in this direction is the
modular LF format introduced in~\cite{rabeEA:twelfmod:09}. This can be converted to \omdoc
by the TWELF system, which makes its treatment directly analogous to \stex, this would
provide a way of information sharing among different authoring systems and workflows.

%%% Local Variables: 
%%% mode: latex
%%% TeX-master: "paper"
%%% End: 

% LocalWords:  svnInfo kohlhase svnKeyword concl ednote

\bibliographystyle{alpha}
\bibliography{kwarc}

\end{document}